\RequirePackage{fix-cm}
\documentclass[smallextended]{svjour3}
\makeatletter\if@twocolumn\PassOptionsToPackage{switch}{lineno}\else\fi\makeatother

\usepackage{amsfonts,amssymb,amsbsy,tabulary,amsmath}
\smartqed  
\usepackage{graphicx}
\usepackage{ulem}

\usepackage{url,multirow,morefloats,floatflt,cancel,tfrupee}
\makeatletter

\AtBeginDocument{\@ifpackageloaded{textcomp}{}{\usepackage{textcomp}}}
\makeatother
\usepackage{colortbl}
\usepackage{xcolor}
\usepackage{pifont}
\usepackage[nointegrals]{wasysym}
\makeatletter

\def\mcWidth#1{\csname TY@F#1\endcsname+\tabcolsep}

\def\cAlignHack{\rightskip\@flushglue\leftskip\@flushglue\parindent\z@\parfillskip\z@skip}
\def\rAlignHack{\rightskip\z@skip\leftskip\@flushglue \parindent\z@\parfillskip\z@skip}

\@ifundefined{etal}{}{}

\usepackage{ifxetex}
\ifxetex\else\if@twocolumn\@ifpackageloaded{stfloats}{}{\usepackage{dblfloatfix}}\fi\fi

\AtBeginDocument{
\expandafter\ifx\csname eqalign\endcsname\relax
\def\eqalign#1{\null\vcenter{\def\\{\cr}\openup\jot\m@th
  \ialign{\strut$\displaystyle{##}$\hfil&$\displaystyle{{}##}$\hfil
      \crcr#1\crcr}}\,}
\fi
}

\AtBeginDocument{%
  \@ifpackageloaded{endfloat}%
   {\renewcommand\efloat@iwrite[1]{\immediate\expandafter\protected@write\csname efloat@post#1\endcsname{}}}{\newif\ifefloat@tables}%
}%

\def\BreakURLText#1{\@tfor\brk@tempa:=#1\do{\brk@tempa\hskip0pt}}
\let\lt=<
\let\gt=>
\def\processVert{\ifmmode|\else\textbar\fi}

\@ifundefined{subparagraph}{
\def\subparagraph{\@startsection{paragraph}{5}{2\parindent}{0ex plus 0.1ex minus 0.1ex}%
{0ex}{\normalfont\small\itshape}}%
}{}

\newcommand\role[1]{\unskip}
\newcommand\aucollab[1]{\unskip}
  
\@ifundefined{tsGraphicsScaleX}{\gdef\tsGraphicsScaleX{1}}{}
\@ifundefined{tsGraphicsScaleY}{\gdef\tsGraphicsScaleY{.9}}{}
\def\checkGraphicsWidth{\ifdim\Gin@nat@width>\linewidth
	\tsGraphicsScaleX\linewidth\else\Gin@nat@width\fi}

\def\checkGraphicsHeight{\ifdim\Gin@nat@height>.9\textheight
	\tsGraphicsScaleY\textheight\else\Gin@nat@height\fi}

\def\fixFloatSize#1{}
\let\ts@includegraphics\includegraphics

\def\inlinegraphic[#1]#2{{\edef\@tempa{#1}\edef\baseline@shift{\ifx\@tempa\@empty0\else#1\fi}\edef\tempZ{\the\numexpr(\numexpr(\baseline@shift*\f@size/100))}\protect\raisebox{\tempZ pt}{\ts@includegraphics{#2}}}}

\AtBeginDocument{\def\includegraphics{\@ifnextchar[{\ts@includegraphics}{\ts@includegraphics[width=\checkGraphicsWidth,height=\checkGraphicsHeight,keepaspectratio]}}}

\DeclareMathAlphabet{\mathpzc}{OT1}{pzc}{m}{it}

\def\URL#1#2{\@ifundefined{href}{#2}{\href{#1}{#2}}}

\def\UrlOrds{\do\*\do\-\do\~\do\'\do\"\do\-}%
\g@addto@macro{\UrlBreaks}{\UrlOrds}

\edef\fntEncoding{\f@encoding}

\makeatother

\newif\ifmultipleabstract\multipleabstractfalse%
\usepackage[authoryear]{natbib}

\usepackage[T1]{fontenc}

\makeatletter	

\AtBeginDocument{%
  \@ifpackageloaded{endfloat}%
   {
\renewcommand*\efloat@process[2]{%
  \ef@ifct{#1}{%
    \expandafter\immediate\expandafter\closeout\csname efloat@post#1\endcsname
    \ef@setct{#1}{0}%
    \clearpage                                                         
        
    \efloat@ifflag{#2list}{
      {\normalsize\efloat@listof{#2}}
    }{}%

    \efloat@ifflag{#2head}{%
      \section*{\@nameuse{#2section}}
      \suppressfloats[t]
    }{}

    \markboth                                                          
      {\expandafter\uppercase\expandafter{\csname #2section\endcsname}}
      {\expandafter\uppercase\expandafter{\csname #2section\endcsname}}

    \def\efloat@type{#2}%
    \processdelayedfloat@hook
    \@nameuse{process#2s@hook}%
    \clearpage
    \@input{\jobname.#1}%
  }{}}
   
   }{}%
}%

  \makeatother

\newcommand*\aap{A\&A}

\newcommand*\ao{Appl Opt}

\newcommand*\apjl{ApJ}

\newcommand*\apjs{ApJS}

\newcommand{\ME}{$\text{M}_{\oplus}$}

\makeindex             

\begin{document}

\title{Setting the Stage: Planet formation and Volatile Delivery}
\titlerunning{Planet Formation and Volatile Delivery}
  
\author{Julia Venturini$^{1}$ \and Maria Paula Ronco$^{2,3}$ \and Octavio Miguel Guilera$^{4,2,3}$
  }

\authorrunning{Venturini \and Ronco \and Guilera} 

\institute{ J. Venturini \\ \email{julia.venturini@issibern.ch} }

\date{Received: date / Accepted: date}

\maketitle

\footnotetext[1]{International Space Science Institute. Hallerstrasse 6, 3012, Bern, Switzerland.}
\footnotetext[2]{Instituto de Astrof\'{\i}sica. Pontificia Universidad Cat\'olica de Chile, Avda. Vicu\~na Mackenna 4860, Santiago 8970117, Chile.}
\footnotetext[3]{N\'ucleo Milenio Formaci\'on Planetaria - NPF, Chile.}
\footnotetext[4]{Instituto de Astrof\'{\i}sica de La Plata (CONICET-UNLP), Paseo del Bosque S/N, La Plata 1900, Argentina.}

\begin{abstract}
The diversity in mass and composition of planetary atmospheres stems from the different building blocks present in protoplanetary discs and from the different physical and chemical processes that these experience during the planetary assembly and evolution. 
This review aims to summarise, in a nutshell, the key concepts and processes operating during planet formation, with a focus on the delivery of volatiles to the inner regions of the planetary system. 
\end{abstract}

\section{Protoplanetary discs: the birthplaces of planets}
\label{Sect:Intro}

Planets are formed as a byproduct of star formation. In star forming regions like the Orion Nebula or the Taurus Molecular Cloud, many discs are observed around young stars \citep{Isella2009, Andrews10, Andrews2018, Cieza2019}.
Discs form around new born stars as a natural consequence of the collapse of the molecular cloud, to conserve angular momentum. As in the interstellar medium, it is generally assumed that they contain typically 1\% of their mass in the form of rocky or icy grains, known as \textit{dust}; and 99\% in the form of gas, which is basically H$_2$ and He \citep[see, e.g.,][]{Armitage10}. However, the dust-to-gas ratios are usually higher in discs \citep{Ansdell2016}. There is strong observational evidence supporting the fact that planets form within those discs \citep{Bae2017, Dong2018, Teague18, Perez2019}, which are accordingly called \textit{protoplanetary discs}. In particular, two young forming planets have been detected recently around the young star PDS 70 \citep{Keppler2018, Muller2018, Haffert2019}.

Protoplanetary discs present a range of lifetimes. This has been inferred, observationally, by two means.
The oldest is the continuum emission of young stellar objects, whose excess in the infrared indicates the presence of warm dust \citep{Lada84}. 
The sharp decay of this infrared emission with stellar age made it possible to estimate that protoplanetary discs last typically between 1 and 10 Myr \citep{Mamajek09,Ribas15}. The lifetime distributions follow an exponential decay, with typical e-folding times of $\sim$3-4 Myr \citep{Mamajek09,Pfalzner2014}. Since dust is very well coupled to the gas, these infrared observations pointed, indirectly, to the dispersal of gas in protoplanetary discs.
A caveat with this deduction is that the detectable infrared radiation is emitted by grains of maximum size of $\sim$cm \citep{Testi2003,Rodmann2006,Ricci2010,Testi2014}. Consequently, the  detected infrared excess could be interpreted, simply, as dust growth up to those sizes within the observed ages. Here is where the second type of discs' observations plays a crucial role: the footprints of the accretion onto the central star. This accretion dissipates energy in very short wavelengths, which is detected in the star's UV spectrum \citep{Gullbring2000}. The accretion onto the central star has typical values of $\sim 10^{-8} \, M_{\odot}$/yr. Mean discs' masses inferred from observations show values of $\sim0.01\, M_{\odot}$ \citep{AW05,Trapman17}. This, together with the values of the accretion rate onto the star give a dispersal timescale of the order of $\sim 10^6$ years, in agreement with the order of magnitude dissipation time inferred from the observations of the warm dust emission.
For the Solar System, thanks to meteoritic records, further constraints can be drawn for the lifetime of the protosolar disc.
Recent measurements of isotopic ages and paleomagnetic analyses of chondrites suggest that the solar nebula was probably gone after $\sim$ 3.8-4.5 Myr \citep{Wang2017, Bollard2017}, in agreement with what is inferred from protoplanetary discs observations.

Different processes, many of them poorly understood, contribute to the disc's dispersal, namely, viscous turbulence \citep{Pringle1981}, disc winds \citep{Suzuki2016}, and photoevaporation \citep{Owen2012}. 
Only since very recently, observations of molecules such as CO, $^{13}$CO, CN and CS with ALMA are starting to constrain turbulence on discs \citep{Teague16}. 
This will shed light on the physical processes that drive disc dissipation. 
\section{Planet formation theory}
\label{Sect:CAM}
Planets that accrete significant amounts of gas, like ice and gas giants, must form before the protoplanetary disc dissapears. As explained above, this is expected to happen in a few million years, which poses a very strong constraint on planet formation models.
There are two broad models to explain how planets form. One is a top-down model, meaning that the formation of massive objects occurs first. This model is known as 
\textit{gravitational instability} \citep{Kuiper1951, Boss1997}, and states that planets form from the collapse of gas into clumps by self-gravity. For this to occur, the clumps in the disc must cool fast in order for gravity to overcome gas pressure. This condition can preferentially be fulfilled at large distances from the star \citep[e.g.,][]{Boss1998,Boss2002,Rafikov2005}. Simulations show that the objects formed by this process are typically of several Jupiter masses \citep[e.g.,][]{Vorobyov2013}, closer to brown-dwarfs or low-mass stellar companions \citep[see][and references therein]{Kratter2016}. The formation of smaller objects is more challenging and requires mass-loss mechanisms, like \textit{tidal-downsizing} \citep{Nayakshin2010, Vorobyov2018}. The other paradigm to understand planet formation is the \textit{core accretion} model. Core accretion has been more extensively implemented than disc instability and it is quite successful to explain the formation of planets with a broad range of masses, orbital parameters and compositions; comparable to what observations show \citep{IdaLin2004a, Alibert05, Benz2014, Ronco2017, Bitsch2017, Mordasini2018}. In what follows, we focus only on core accretion. 

\subsection{The core accretion model}
Planets form embedded in protoplanetary discs. The farther from the star, the lower the temperature. At different temperatures, different materials are able to condense. 
In core accretion, these condensates are generically called \textit{solids} or \textit{heavy elements}. 
The central idea of the core accretion model is that a planet forms first by the accumulation of solids or heavy elements into a core, followed by the binding of an atmosphere on top \citep{Safronov1969,mizuno78, BP86,P96}. Since the dominant gas constituent of the disc is H$_2$ and He, this is usually considered the main composition of primordial atmospheres.
The atmosphere or envelope starts to form when the gravity at the core's surface overcomes the local gas sound speed, typically at approximately a lunar mass \citep[but depends on the distance to the star, see][]{Armitage10}. This means that very early on during the formation, a protoplanet grows by accreting both solids and gas.

Initially, the solid accretion rate is higher than the gas accretion rate, so the core grows faster than the gaseous envelope. However, when the mass of the atmosphere or envelope becomes comparable to the mass of the core, the envelope starts to be compressed efficiently by its self-gravity. This allows more gas to enter the protoplanet's gravitational sphere of influence, which enhances even more the envelope's self gravity.  As a consequence, a runaway gas accretion process is triggered. When this happens, the core is said to have reached the \textit{critical core mass}. In classical models, the critical core mass ranges in value between $\sim$5 and 15 Earth masses (hereafter \ME), depending on the envelope's opacity, solid accretion rate and planet location \citep{mizuno80, Ikoma00}.

\begin{figure}
  \centering
  \includegraphics[width=\textwidth]{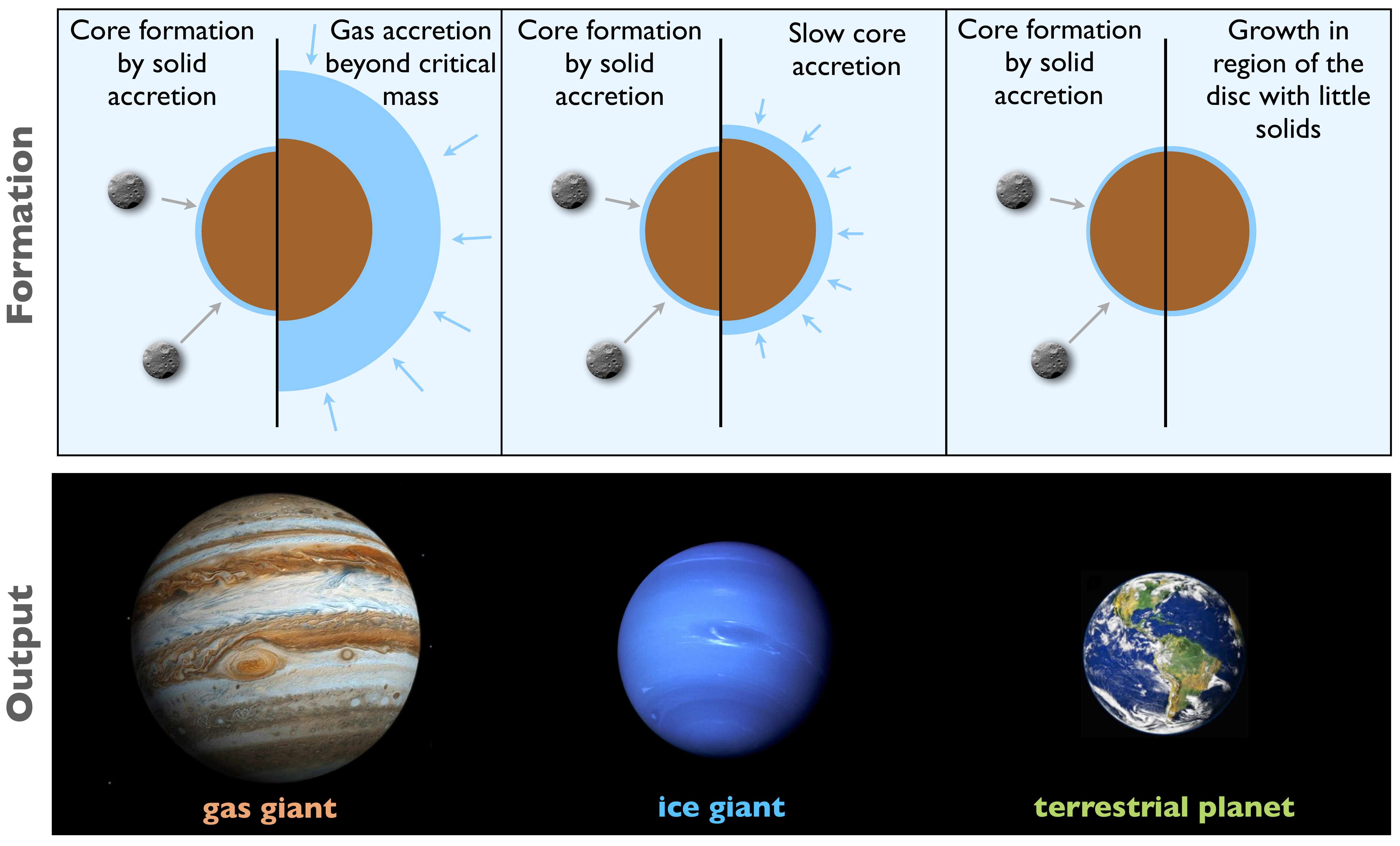}
  \caption{Possible conditions and outcomes of core accretion to explain Solar System planets. Courtesy of Y. Alibert (adapted for this review).}
  \label{fig_CA}
\end{figure}

Despite that core accretion was originally developed to explain the formation of gas giants like Jupiter \citep{PerriCam74,mizuno78,mizuno80,BP86}; this model has become the paradigm of planet formation for its ability to account for a large diversity of planetary outcomes \citep[see, e.g,][]{Benz2014}.   
Regarding the Solar System planets, in the view of core accretion, a gas giant like Jupiter can form if the core reaches a critical mass when there is still plenty of gas in the disc. In this way, runaway of gas can occur, allowing for the accretion of hundreds of Earth masses of gas. 
Alternatively, if the embryo grows in a region of the disc where solid accretion is slower, it could happen that by the time of reaching the critical mass, the gas in the disc is scarce. Hence, even if the core is critical, gas accretion by large amounts will not occur. In this case we would end up with a core-dominated planet like Uranus and Neptune, which have a significant H$_2$-He envelope \citep[$\sim$10 -25 \% of the planetary mass, ][]{Helled11} but where the gas is not the dominant constituent.
Finally, if the solid accretion timescales are much larger than the disc dispersal timescales, or if the embryo grows in a region of the disc where there are not many solids to accrete (e.g., close to the central star), it will remain basically as an embryo during all the disc's lifetime \citep{Fortier13}, binding, in some cases, a thin primordial hydrogen-dominated atmosphere.\footnote{If the embryo reaches a mass of $\gtrsim 0.5-0.8$ \ME \, while still embedded in the gas disc, it can bind a primordial atmosphere of $\sim 10^{-4}-10^{-2}$\ME \, \citep{Lee15}. The primordial atmosphere can be lost by \textit{boil-off} in the $\sim 10^5$ years following disc's dispersal, depending on the planet's mass and proximity to the star \citep{Owen2016}.}  
After the disc dissipates, dynamical instabilities occur, leading to collisions of embryos via giant impacts. This is how Earth formation is envisioned, since the very early works on the subject \citep[see, e.g,][]{Wetherill85, Wetherill90}. More recent literature on terrestrial planet formation is summarised in \citet{Izidoro18} and \citet{OBrien18}, and explained briefly in Sect.\ref{Sect:Volatiles2}. The classical picture of core accretion and its output in terms of Solar System planets is summarised in Fig. \ref{fig_CA}.

It is important to mention that the formation of intermediate-mass planets, like our ice giants, is not completely understood. The problem with the scenario described above is that it involves fine-tuning: the dispersal of the disc has to occur when the protoplanet is $\sim$10-20 \ME, mass range at which gas accretion is expected to be extremely fast. Thus, stopping gas accretion at that mass is unlikely \citep{Venturini17}, making the formation of intermediate-mass planets very rare, in contradiction with observations \citep{Batalha13, Suzuki18}. 

Another important caveat is that classical models assume for simplification that the heavy elements and the gas do not mix: the solids always reach the core while the hydrogen and helium remain on the outermost layer, constituting the primordial atmosphere \citep{BP86, Ikoma00}. This is of course not what is expected to happen. In reality, the incoming solids get heated by gas drag while crossing the atmosphere. Ice sublimation is expected, together with mechanical break up \citep{Podolak88, Mordasini06, Iaroslav07, Claudio19}. As a consequence, material originally in solid state can mix in vapour form with the  preexisting hydrogen and helium \citep{Brouwers2018}. When this effect is included in formation models, simulations show that due to the increase in mean molecular weight \citep{Venturini16}, and due to the reduction of the adiabatic gradient by chemical reactions \citep{HI11, Venturini15}, the compression of the envelope due to self gravity occurs more effectively, for smaller cores. Consequently, gas giants can form faster \citep{Venturini16}, and small-mass planets with substantial H-He atmospheres ($\sim$ 10-20\% of planetary mass, the so-called \textit{mini-Neptunes}) can be expected as well \citep{Venturini17}. 

\subsection{Dominants size of the accreting solids: planetesimals and/or pebbles?}
\label{Sect:Sizes}

Core accretion can only take place if a primordial embryo or core acts as a seed for the growth. How is this seed produced?
Dust particles grow and settle to the midplane of discs in very short timescales ($\sim10^3$ yr), until reaching centimetre-size  \citep{Weiden77, Dullemond05, Brauer08}.
These \textit{pebbles} are partially coupled to the gas, meaning that the gas dynamics strongly affects pebbles' orbits. From the equation of motion of a gas particle embedded in a protoplanetary disc, one can find that the azimutal velocity of gas particles satisfies \citep{Weiden77}: 
\begin{equation}\label{gas_vel}
v_{g,\theta}^2 = v_k^2 + \frac{r}{\rho} \frac{\partial P}{\partial r}
\end{equation}
where $v_{g,\theta}$ is the azimutal gas velocity, $v_k$ the keplerian velocity (the azimutal velocity of the solid particles, which are not pressure-supported), $r$ the radial distance to the star, $\rho$ and $P$ the local gas density and pressure, respectively. In a standard smooth disc, the pressure decreases outwards, which makes the second term of Eq.\ref{gas_vel} negative. As a consequence, the gas moves at a slower rate than the pebbles. Hence, from the pebble's reference frame, the gas acts like a headwind. This provokes a strong orbital decay towards the central star. For instance, a 1-meter object at 1 AU would drift towards the central star in only $\sim$ 100 years \citep{Weiden77}.
Because of this radial drift barrier, the growth of dust aggregates larger than centimetre/meter has been a puzzle for planet formation theory for decades \citep[see, e.g.][and references therein]{Blum2018}. At the same time, other growth barriers exist. Bouncing, fragmentation and erosion of dust aggregates are the general outcome of collisional laboratory experiments when the size of the colliding bodies is similar, and mass transfer and cratering are the most common ones among different size objects \citep[see, e.g.][]{Guttler2010,Birnstiel2016}.

Nowadays, the most successful theory capable of closing the gap between dust growth and the formation of km-size objects (called \textit{planetesimals}) is the \textit{streaming instability} \citep{Youdin05,Johansen07}. This instability comes from a back-reaction of the dust to the gas: the gas in the vicinity of the dust particles is accelerated, which decreases locally the gas friction into the dust. Hence, dust particles move slower towards the star, piling up particles coming from farther out. These pile ups or clumps of dust eventually collapse into planetesimals by self-gravity \citep{JohansenPPVI}. 
The direct collapse allows to surpass the aforementioned growth barriers, favouring the
formation of big planetesimals ($\sim$10-100 km). 
Some caveats with the streaming instability is that it requires certain fine-tuned conditions to occur, like a high dust-to-gas ratio and low viscosity \citep{Youdin05}. Moreover, \citet{Krapp2019} showed that if a size distribution of dust particles is considered, the streaming instability acts on longer time scales than expected if a single dust size is used. On the other hand, the formation of sub-kilometers planetesimals by direct sticking cannot be completely discarded \citep[see][and references therein]{Blum2018}.

A key unknown when modelling planet formation is the dominant size of the solids at the time when planets assemble. The first planet formation models assumed the dominant size to be 100~km planetesimals \citep{P96}. That assumption originates from the presence of these objects in the Solar System, thought to be left overs of planet formation \citep{Safronov1969, Whipple1972}. 
Indeed, the largest asteroids in the Solar System are in the 100~km size range, and many smaller bodies are known to have formed via high-velocity collisions \citep{Bottke2005}. 
However, the initial dominant size of planetesimals is still under debate. \citet{Morby09} showed that the actual size distribution of the asteroid belt can be reproduced from initially big planetesimals (100-1000~km), but \citet{Weidenschilling2011} showed that such distribution can also be obtained from initially small planetesimals ($\sim$100~m). \citet{Kenyon2012} found that the actual size distribution of the trans-Neptunian objects is better reproduced from initially $\sim$1-10~km planetesimals. Indeed, some Kuiper Belt Objects like Pluto, Charon and MU69, do not show crater impacts with sizes below $\sim$1 km \citep{Stern2019}. 
\citet{SchreiberKlahr18} showed, via streaming instability, that planetesimals should be of the order of 100~km size in the inner disc but could present a variety of sizes in the outer one. Thus, planetesimals could in fact present different dominant sizes at different orbital distances from the star.

A well-known problem of growing planets by the accretion of large planetesimals is that the timescale to form a gas giant tends to be too long\footnote{Nevertheless, \citet{Guilera10, Guilera11} showed that the formation timescales are strongly reduced if giant planets form by the accretion of sub-km planetesimals, and that the simultaneous formation of Solar System giant planets can occur in only a few Myr (compatible with disc lifetimes). The prevalence of such small planetesimals at the time of planet formation is however not predicted by streaming instability simulations. \citep{SchreiberKlahr18}}. This happens because planetesimals get excited by gravitational interactions between embryos and planetesimals, which increases the relative velocities among them. Planetesimals (especially big ones) are not well coupled to the gas, so the damping of eccentricity and inclination by gas friction operates poorly on 100-km size objects \citep{Fortier07,Guilera10,Fortier13}. As a consequence, relative velocities between big planetesimals and embryos are high, making big planetesimals hard to accrete. Under these circumstances, planetesimals halt their growth and embryos are the only bodies that continue growing at a slow pace (although faster than in orderly growth). This type of solid accretion is known as \textit{oligarchic growth} \citep{IdaMak93,KI98}. 
To match observations, planetesimal-based models accounting for oligarchic growth require a planetesimal size of 300 m \citep{Fortier13,Alibert2013,Mordasini2018}. This is not the dominant size of planetesimals predicted by streaming instability. Actually, in situ simulations can lead to the formation of a gas giant when 10-100 km size planetesimals are considered, as long as disc masses and/or metallicities are high \citep{Fortier09, Guilera14}. However, when migration is included, because migration timescales are shorter than the core growth timescales with large planetesimals, planets are typically lost into the star before reaching critical masses \citep{Fortier13,Ronco2017}. The choice of 300 m size makes the migration and core growth timescales comparable, and thus the formation of gas giants becomes possible in metal-rich discs \citep{Fortier13}. Hence, the problem of planetesimal accretion should not be viewed apart from the problem of the persistent too fast migration rates.

On the contrary, if one assumes that the dominant solids in the disc at the time of planet formation are pebbles, because of the strong orbital decay mentioned above, the growing embryo sees a flux of pebbles passing through its orbit. Since pebbles are effectively slowed down by gas friction, they fall readily towards the embryo once they enter its gravitational sphere of influence, making them much easier to accrete than planetesimals \citep{OrmelK10, JohansenLacerda10,JohansenLambrechts2017}. Consequently, a critical core can be made fast by pebble accretion; for example, in only 1 million years at a distance of 5 AU from a solar-type star \citep{LJ14}, which made pebble-based models very attractive.
However, gas giants with periods shorter than ten years are expected to be less than $\sim 15$\% of the existing exoplanets \citep{Mayor2011, Howard2012, Hsu2018}.
Thus, 
the easiness of pebble accretion to produce gas giants seems to be in conflict with observations \citep{Ndugu2018}. This can be overcome if a late formation time for the embryos is assumed \citep{Brugger18} or if a high irradiation environment like a stellar cluster is invoked \citep{Ndugu2018}.
We note, nevertheless, that pebble accretion is still a relatively recent and constantly evolving scenario. In particular, since the pebble drift and accretion are so closely connected to the disc properties, testing different and improved disc models is essential to make thoroughly comparisons with observations. Other physical aspects like envelope enrichment \citep{Venturini17} and gas recycling \citep{Ormel15} might also play an important role on the gas budget of planets formed by pebbles.

\subsubsection{Isolation Mass}
\label{Sect:Iso}
 An important concept in both pebble and planetesimal accretion is that of the \textit{isolation mass}, although it has different meanings in the different contexts.
 In pebble accretion, there is a certain time in the planet's growth when the protoplanet perturbs the disc enough to create a pressure bump beyond its orbit \citep{Lambrechts14}. 
 Within a pressure bump, the pressure gradient of the disc (second term of the right hand side of Eq.~\ref{gas_vel}) becomes positive. Hence, the pebbles in that location feel a tailwind that makes them spiral outwards until the pressure maximum.
 Beyond the maximum, the solid particles tend to move inwards as usual. Consequently, the pressure bump acts like a particle trap \citep{HB2003}, and the pebbles that fall within it cannot reach the protoplanet. 
 Once this mass is reached, the protoplanet is said to have reached the \textit{pebble isolation mass} \citep{Lambrechts14,Sareh18,Bitsch18}, and pebble accretion stops. 
On the other hand, planetesimals are accreted within the protoplanet's \textit{feeding zone}. This is the region adjacent to the embryo where the embryo dominates gravitationally, typically an annulus of $\sim$10 Hill radius centred in the protoplanet \citep{TanakaIda99}.
The \textit{planetesimal isolation mass} is reached when the embryo's feeding zone runs out of planetesimals. Despite of this, planetesimal accretion does not cease completely.
Since the tenuous envelope surrounding the core keeps contracting and accreting gas, the gravitational influence of the protoplanet keeps expanding. 
Hence, new planetesimals enter slowly in the practically empty feeding zone, making it possible for planetesimal accretion to continue \citep[the so-called phase 2 of][]{P96}. 
Note, however, that phase 2 does not always take place when a planet grows by planetesimals. When the gravitational excitation from the embryo into the planetesimals is taken into account,
the initial phase of core growth can be very slow, meaning that the feeding zone never gets depleted. Indeed, at a distance from the star larger than a few AU, phase 2 is unlikely to exist \citep{Fortier07,SI08,Guilera10,Venturini20}.

\subsubsection{Open problems with planetesimal and pebble accretion rates}
\label{Sect:discussion_solid_acc}

Global models based on planetesimal accretion rely on prescriptions for the accretion rates derived from statistical collisions from N-body simulations and in situ accretion \citep[e.g.][]{Inaba2001}. These accretion rates can be reduced by the planetesimal trapping mechanism \citep{TanakaIda97}. In this scenario, the planetesimals that are scattered by the protoplanet are then damped by the gas drag and trapped outside the planet's feeding zone. However, when several embryos grow in the disc, the scattering of planetesimals from one planet's feeding zone can replenish the feeding zone of others \citep{TanakaIda97}. 
Furthermore, \citet{TanakaIda99} showed that a planet migrating through a swarm of planetesimals can acts as a shepherd or as a predator, reducing or increasing, respectively, the accretion rates.
Recently, \citet{Shibata2020} showed that planetesimal accretion can play an important role in the late enrichment of migrating giant proto-planets due to dynamical effects that break the planetesimal trapping. In summary, planetesimal accretion is a very complex process that requires further revision. 

Planet formation models based on pebble accretion also suffer from certain simplifications at the moment of computing accretion rates. For instance, pebble accretion rates depend directly on the Stokes number and pebble surface density (or pebble flux) at the position of the planet \citep{Lambrechts14}, which are often taken as constant or evolving in time following an exponential decay \citep{Lambrechts19, Ogihara2020}. However, \citet{Drazkowska16,Drazkowska2017} have shown via dust evolution and pebble formation models that the Stokes number and pebble flux change in time and along the disk in a more complex way. In addition, most of the models assume an infinite pebble supply \citep{LJ14, Lambrechts14,Bitsch19}, when in reality pebbles originate from a finite reservoir of dust and are lost towards the star by drift and diffusion \citep{Birnstiel12, Drazkowska16}. Therefore, future pebble accretion models should address the fundamental and unavoidable interconnection between disc evolution, and dust growth, drift and fragmentation.

\subsection{Planet migration} 
\label{Sect:Mig}

As a planetary embryo grows, the gravitational interaction between the planet and the gaseous disc produces torques that modify the planet's orbit, making the protoplanet migrate along the disc \citep{GT1979,LP1986}. There are two main migration regimes. The first one, known as type I migration, involves planets that are not massive enough to open a gap in the gaseous disc \citep{Ward1997}. In this regime, the gravitational interactions between the planet and the disc become significant at the Lindblad resonances and at the corotation region of the planet's orbit. In addition, when the mass of the planet is small enough to allow that planet-disk interactions to be treated using linear approximations, the migration rate is  proportional to the planet's mass and to the gas surface density. The second regime, the type II migration, involves massive planets where the gravitational interactions are strong enough to open a gap in the gaseous disc around the planet's orbit \citep{LP1986}. Such planets are typically more massive that Saturn, but the exact transition value between type I and type II migration depends on the disc's parameters \citep{Crida2006}. Planet migration is key in the orbital evolution of forming planets, having a strong impact on their final masses and semi-major axes.

\subsubsection{Type I migration}
If $\Gamma$ is the total torque over the planet in the type I migration regime, the change of the semi-major axis of the planet, $a_{\text{P}}$, is given by
\begin{equation}
  \frac{d a_{\text{P}}}{dt}= \frac{2 a_{\text{P}} \Gamma}{\mathcal{L}_{\text{P}}},
    \label{eq:typeI_migration}
\end{equation}
where $\mathcal{L}_{\text{P}}$ is the angular momentum of the planet. The total torque is given by the sum of the two main contributions:
\begin{equation}
  \Gamma= \Gamma_{\text{Lindblad}} + \Gamma_{\text{corotation}},
  \label{eq:torques}
\end{equation}
the Lindblad torque, $\Gamma_{\text{Lindblad}}$, and the cororation torque, $\Gamma_{\text{corotation}}$. The first torque arises from the gravitational interactions between the planet and the disc at the Lindblad resonances. These occur in the locations of the disc where the ratio between the angular velocity of the gas, $\Omega$, and that of the planet, $\Omega_{\text{P}}$, are related by $\Omega / \Omega_{\text{P}} \simeq N / (N \pm 1)$, with $N$ an integer. At the Lindblad resonances, density waves that transport angular momentum away from the planet are launched. In typical protoplanetary discs, the outer Lindblad resonances lie closer to the planet than the inner ones. This makes the negative torque exerted to the planet by the outer Lindblad resonance to be stronger than the positive torques generated by the inner ones.
As a result, a net negative torque from the Lindblad resonances is the rule.   
On the other hand, the corotation torque arises from the gravitational interactions in the co-orbital region of the planet, where $\Omega \simeq \Omega_{\text{P}}$, and the gas presents horseshoe-type orbits. Due to the imbalance in the torques generated by the gas particles moving in the horseshoe orbits, a net positive torque is generated. This happens because gas particles beyond the planet's orbit with higher angular momentum and with lower temperature switch to an inner orbit, respect to the planet's orbit, with less angular momentum and where the gaseous disc is hotter, and viceversa. 

Early hydrodynamical models of type I migration were developed for isothermal discs \citep{Tanaka02}. For typical protoplanetary discs in which the gas surface density decreases with orbital distance, these models found high inward type I migration rates. Analytical prescriptions for these migration rates, derived from the hydrodynamical simulations, were incorporated by many authors to study the role of planet migration in growing planets \citep{IdaLin04,Alibert05,Mord09, Miguel2011, Ronco2017,Miguel2020}.
Since isothermal migration rates are so high, all these previous authors had to reduce ad-hoc such rates by a factor of up to 100--1000 to reproduce observations (especially the mass versus semi-major axis diagram of the exoplanets). This result made it clear that a deeper understanding of migration was needed. A first step towards this direction was to study corotation torques more in depth. \citet{Masset2006} found that corotation torques can abate migration significantly, and even reverse it. This can happen in discs with shallow surface density profiles or in regions where the local radial gradient of the gas surface density becomes positive, like in a disc cavity or in a pressure maximum. Because of this diversity of disc structures, zero torque locations can appear in the discs \citep{Masset2006,Romanova2019}. Another effect that causes positive radial gradients in the gas disc is the presence of a gap produced by photoevaporation due to the central star. This phenomenon, synchronized with type I migration, can also produce outward migration traps \citep{Guilera2017b}. \citet{Guilera2017} also showed that pressure maxima gererated at the edges of a dead zone can act as planet migration traps. These planet traps can be preferential locations for the formation of massive cores not only because planet migration is halted, but also because pressure maxima stop the radial drift of the pebbles and planetesimals, allowing for the accumulation of solid material.

A second important improvement in modelling planetary migration was to account for radiative transfer in hydrodynamical simulations  \citep{PM2006}. This also showed that type I migration can be slowed down, and even reversed in typical protoplanetary discs. These simulations found that the outward type I migration strongly depends on the planetary mass, semi-major axis and disc thermodynamic \citep{Kley2009, BK2011, Paardekooper2011, jm2017}. For non-isothermal discs, \citet{Paardekooper2011} and \citet{jm2017} derived analytical recipes for type I migration rates based on the results of the hydrodynamical simulations.\footnote{While \citet{Paardekooper2011} performed 2D hydrodynamical simulations, the simulations by \citet{jm2017} were performed in 3D. The main difference between the derived recipes from both authors relies in the horseshoe drag component of the corotation torque \citep{Guilera2019}.} One important result found in planet formation models when considering the evolution of non-isothermal discs and the corresponding type I migration rates is the presence of \textit{convergent zones} in protoplanetary discs wherein planet migration is halted \citep{Lyra2010, Cossou2013, Bitsch2013, Dittkrist2014,Bitsch15b,Baillie2016}. Due to the dependence of type I migration with the disc thermodynamics and with the planet mass, the convergent zones change in time as the disc evolves.

Additionally, \citet{Benitez-llambay2015} showed that if the heat due to solid accretion is included in the thermal budget of the nearby disc, two asymmetric hot- and low-density lobes appear, producing a new positive component in the total torque over the planet, known as \textit{heating torque}. On the other hand, \citet{Lega2014} found from radiative hydrodynamical simulations that around a low-mass planet the gas is cooler and denser than in the adiabatic case. This effect is also asymmetric and generates a negative torque, called \textit{cold torque}. Based on these results, \citet{masset2017} developed analytical prescriptions to estimate the total torque that arises from combining the two effects, referred as \textit{thermal torque}. These analytical prescriptions can be incorporated in planet formation models. Recently, \citet{Guilera2019} showed that the inclusion of the thermal torque can generate a significant outward migration on low-mass growing planets if solid accretion rates are high enough. 
We note, however, that \citet{Chrenko2019} showed through 3D radiative hydrodynamic simulations that the linear perturbation theory developed by \citet{masset2017} cannot provide a full description of the heating torque when non-uniform opacities in the disc are considered.

The effect of including the different type I migration prescriptions in planetary growth simulations is illustrated in Fig. 2, where we show growth tracks in mass and semi-major axis of a planet starting with the same conditions under the different prescriptions. In general, the appearance of outwards migration gives the protoplanet more time to grow before reaching the inner regions of the disc. For these test cases, the inclusion of the heating torque allows the planet to grow large enough to open a gap and transition to type II migration. 
 
 \subsubsection{Type II migration}
When a planet becomes massive enough to open a gap in the disc, the planet-disc interaction changes significantly. In the idealised case, the planet is locked inside the gap (it does not interact with the gap nor does it move with respect to it) and the planet migrates together with the gap as the disc evolves \citep{LP1986}. However, when the planet mass becomes comparable to the local disc mass, the interaction between the planet and the gap edges cannot be neglected and the migration rate slows down \citep{Armitage2007}. Moreover, if accretion onto the planet through the gap is considered, type II migration can deviate significantly from the idealised case \citep{DK2015}. More recently, \citet{HP2018} showed that the irradiation from the central star onto the outer gap edge can decrease --and even reverse in extreme cases-- the type II migration of giant planets, allowing the survival of giant planets at moderate distance from the central star. On the other hand, if two planets are massive enough and relatively close to each other to open a common gap, the type II migration of the system can be very different with respect to the idealised case mentioned above. For example, \citet{MassetSnellgrove2001} showed that if the outer planet is less massive, the migration of both planets can be outwards. This is because the more massive inner planet suffers a greater positive torque from the inner disc than the negative torque that suffers the outer planet from the outer disc. Therefore, the total torque over the system of planets can become positive. 

\begin{figure}
  \centering
  \includegraphics[width=\textwidth]{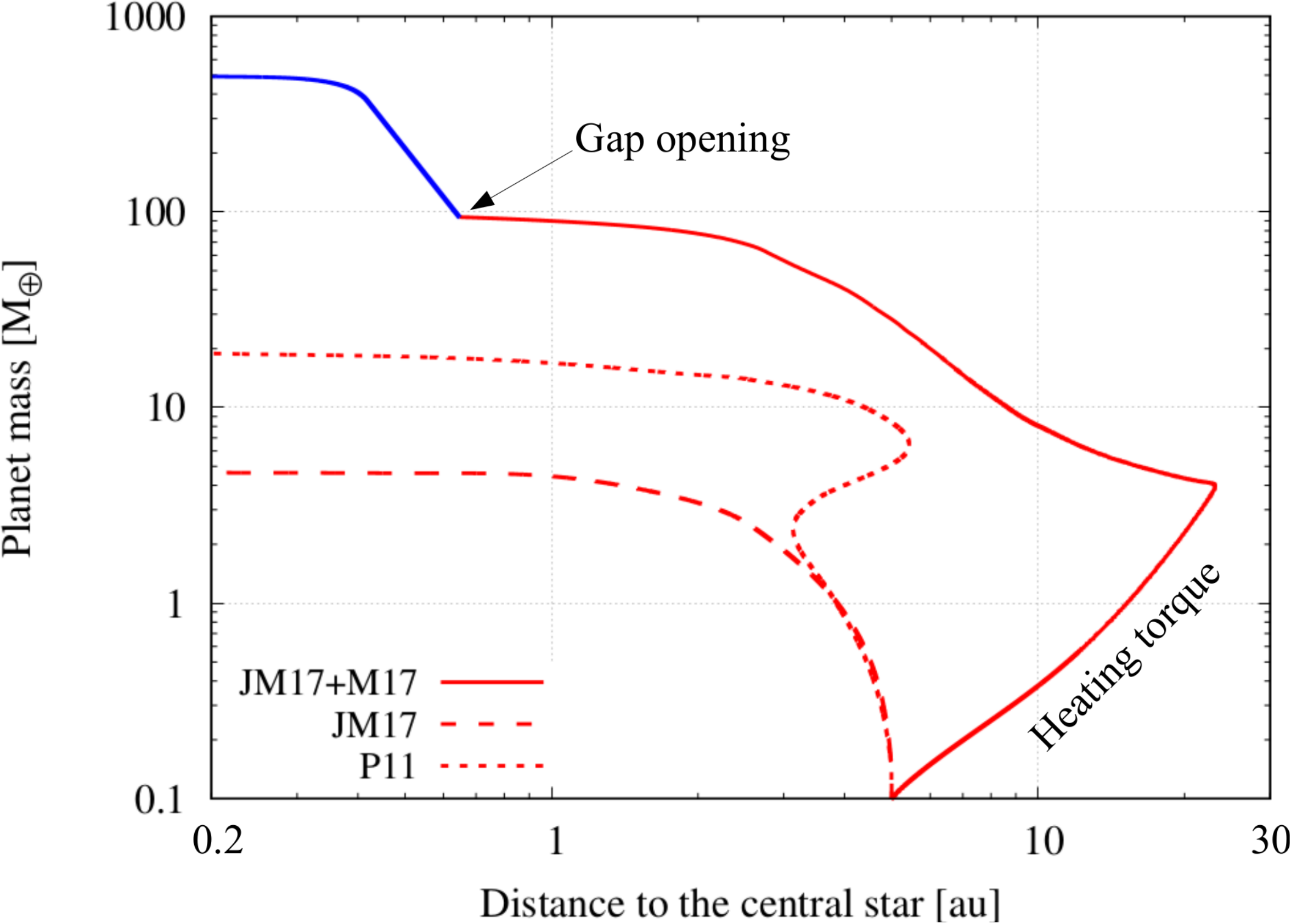} 
  \caption{Planet formation tracks for an initial Mars-mass embryo, initially located at 5~au, that grows by the concurrent accretion of gas and of 300~m-planetesimals (typical size adopted in planetesimal-based simulations, see Sect.\ref{Sect:Sizes}). The embryo is immersed in a disc five times more massive than the Minimum Mass Solar Nebula \citep[][]{Hayashi81}. This corresponds to an initial planetesimal surface density of 
  $\sim 13$ g~cm$^{-2}$ at 5 au.
  The red (blue) color represents the type I (type II) migration regime. For the dotted line, type I migration prescriptions from \citet{Paardekooper2011} are used, while for the dashed curve the recipes adopted are those of \citet{jm2017}. In the first case, the planet reaches an outward migration region with a mass of $\sim 2.5~\text{M}_{\oplus}$. The outward migration is reversed when the planet reaches a mass of $\sim 6~\text{M}_{\oplus}$. For the solid line, type I migration recipes from \citet{jm2017} and the thermal torque from \citet{masset2017} are considered. In this case, the heating torque produces a significant outward migration. This situation allows the planet to have more time to grow, to open a gap and to switch to type II migration. All the formation tracks were calculated using the planet formation code P{\scriptsize LANETA}LP \citep{Guilera2019}. Simulations end when the planets reach 0.2 au (at $\sim 1$~Myr when the type I migration recipes from \citet{Paardekooper2011} and \citet{jm2017} are used, and at $\sim 2$~Myr when the thermal torque is also considered).}
  \label{fig:formation_tracks}
\end{figure}

\section{Volatile enrichment in the inner regions of planetary systems}
\label{Sect:Volatiles}
In planet formation, \textit{volatiles} denote all compounds with a condensation temperature less than that of water.  
Together with this concept, another important one is that of \textit{ice-lines}: locations in the protoplanetary disc where the temperature is low enough for a certain compound to condense. 
There are several ice-lines in protoplanetary discs, the water ice-line being the most famous, and occurring at T$\sim$ 170 K. Very often the water ice-line is simple referred as \textit{the} ice-line.
Since discs cool down with time, the ice-lines move inwards. Nevertheless, this does not necessarily mean that the place where we can find ice moves inwards, this depends on the dominant size of the solids.
If the dominant solids are $\sim$10-100 km-size planetesimals, these will practically not drift along the disc. Hence, the place where ice condenses initially is the place where ice will remain; 
the location of T$\sim$ 170 K will move inwards, but it will be dry.
If the dominant solids are cm-size pebbles, the picture is more complex, the location of the ices will depend on the relative velocity between the pebble drift and the movement of the ice-line \citep{Morby16}.
As the disc cools, icy pebbles can reach the inner regions of the system, polluting the planets that are growing there with water \citep{Sato16}.
In principle, this poses a problem for the Solar System, where the inner planets are basically dry. \citet{Morby15} proposed that the growth of Jupiter can solve this issue. We explain this in Sect.\ref{Sect:Jupiter}.
 
Volatiles can be present in planets from the early stages of planet formation, when there is still gas in the disc; or reach the protoplanets after the gas is gone. 
In the former case, in principle only planets located beyond the ice-line can have volatiles, although water could also be produced within a dry protoplanet from a reaction between H$_2$ and FeO \citep{IkomaGenda2006}, or retained in olivine grains via adsorption \citep{Drake05}. 

The case where volatiles arrive late, after or during the disc's dispersal, is the one broadly accepted for the origin of volatiles in the inner Solar System \citep{Morbidelli2000, Raymond2004, OBrien2006, Raymond2009, Walsh11, RaymondIzidoro2017, Ronco2018}. This \textit{late} delivery might also play a role on the volatile distribution within exo-planetary systems, even in those presenting a different architecture compared to the Solar System \citep{Ronco2014,Ronco15,Zain2018,Sanchez2018}.

\subsection{Water on Earth}
\label{Sect:Water}
The water on the Earth crust, usually referred to as an Earth Ocean (EO), was estimated by \citet{Lecuyer1998} to be $\sim2.4\times10^{-4}$ \ME. However, the water content of the mantle and core remain uncertain. 
Estimations suggest that water on the mantle can vary from $\sim$0.3-3 EO \citep{Lecuyer1998} to $\sim 8$ EO \citep{Marty2012}, but the primitive mantle could contain even higher amounts, between $\sim$10 and $\sim$50 EO \citep{Dreibus1989,Abe2000}.
The core's water is also poorly constrained. It could vary between less than $\sim$0.1 EO \citep{Badro2014} and $\sim$80 EO \citep{Nomura2014}. Hence, a maximum amount of water on Earth would be $\sim$ 0.1\% - 0.2\% \ME. 

Could water on Earth have been produced {\it in situ}? As we mentioned above briefly, water could had been adsorpted {\it in situ}, from the primordial nebula, by olivine grains \citep{Stimpfl04, Drake05, Muralidharan2008}. By this mechanism the Earth could have acquired several EO of water. This is in contradiction with studies claiming that the first $\sim$60-70\% of the Earth's mass was assembled by oxygen-poor building blocks \citep{Rubie2011}. Alternatively, water could have been produced in situ by a reaction between the H$_2$ of a primitive atmosphere with the iron-oxides of the magma ocean \citep{IkomaGenda2006}. The authors show that for that reaction to occur, the planet's mass at the time of disc dispersal has to be above $\sim$0.3 \ME. For a less massive embryo, the temperature at the bottom of the atmosphere is below 1500 K, the melting temperature of silicate. Recent studies show that the Earth had a mass of $\approx$ 0.5-0.75 \ME \, at the time of disc dissipation \citep[see ][and references therein]{Lammer2018}\footnote{Indeed, proto-Earth must have grown large enough during the disc phase to bind a solar-composition atmosphere able to account for the present-day noble gases \citep{Marty2012}, but small enough to be able to lose such primordial atmosphere in the course of giga-years of thermal evolution \citep{Lammer2018}.}, meaning that the Ikoma-Genda mechanism could have operated on Earth. In their scenario, once the protosolar disc is gone, the planet cools and the ocean forms by condensation of the steam atmosphere in about a thousand years. 

A major argument typically used to discard an in-situ origin for Earth's water is that the D/H value of the Earth's ocean is roughly $\sim$7 times the proto-solar value \citep{Drake02}. There are however two caveats with this. First, the D/H of the oceans has probably evolved with time, increasing by a factor of $\sim1.5-2$ \citep{Pahlevan19}, or $\sim2-9$ \citep{GendaIkoma08} due to equilibrium partitioning that results from atmospheric escape. Second, the D/H of the oceans might not be representative of the Earth's bulk D/H. Indeed, measurements from lavas (representative of the Earth's mantle, where water could have been kept pristine), point towards a lower D/H ratio, closer to the protosolar value \citep{Hallis2015}. 
This suggests that the hypothesis of nebular water acquisition might account for some part of the Earth's water content. 
Indeed, calculations based on the abundance of terrestrial Ne isotopes suggest that about 10\% of the Earth's water could have originated from a primordial H$_2$-atmosphere \citep{Marty2012}.

From where could the other $\sim$90 \% of the Earth's water come from? If the bulk of water was delivered to the Earth, which kind of primitive objects of the Solar System were the main responsible? Comets were the first considered due to their high ice content \citep{Chyba1987,Delsemme1998}. However, they are generally discarded for two main reasons. First, the comets with measured D/H ratio present values that are higher than that of the Earth by a factor of $\sim$2-4 \citep{Altwegg15}. Second, it is difficult to explain, from a dynamical point of view, the delivery of the amount of water on Earth only through cometary impacts. The most optimistic scenario points to a maximum of $\sim$10\% of the Earth's water originating from this population \citep{Morbidelli2000}. Additional evidence from noble gases shows that the cometary contribution should be less than $\sim$1\% \citep{Marty2012}. Moreover, \citet{Marty2012} shows that the isotopic composition of hydrogen and nitrogen of Earth's surface corresponds to the one of carbonaceous chondrites, suggesting that planetesimals and/or embryos from the outer asteroid belt were the main source of Earth's volatiles. Although new models including results from the Rosetta mission still support that the cometary contribution to the Earth’s water inventory was little ($\le 1$\%), they suggest it may have been significant for the supply of noble gases \citep{Marty2016}. 

\begin{figure}
 \centering
 \includegraphics[width=0.8\textwidth]{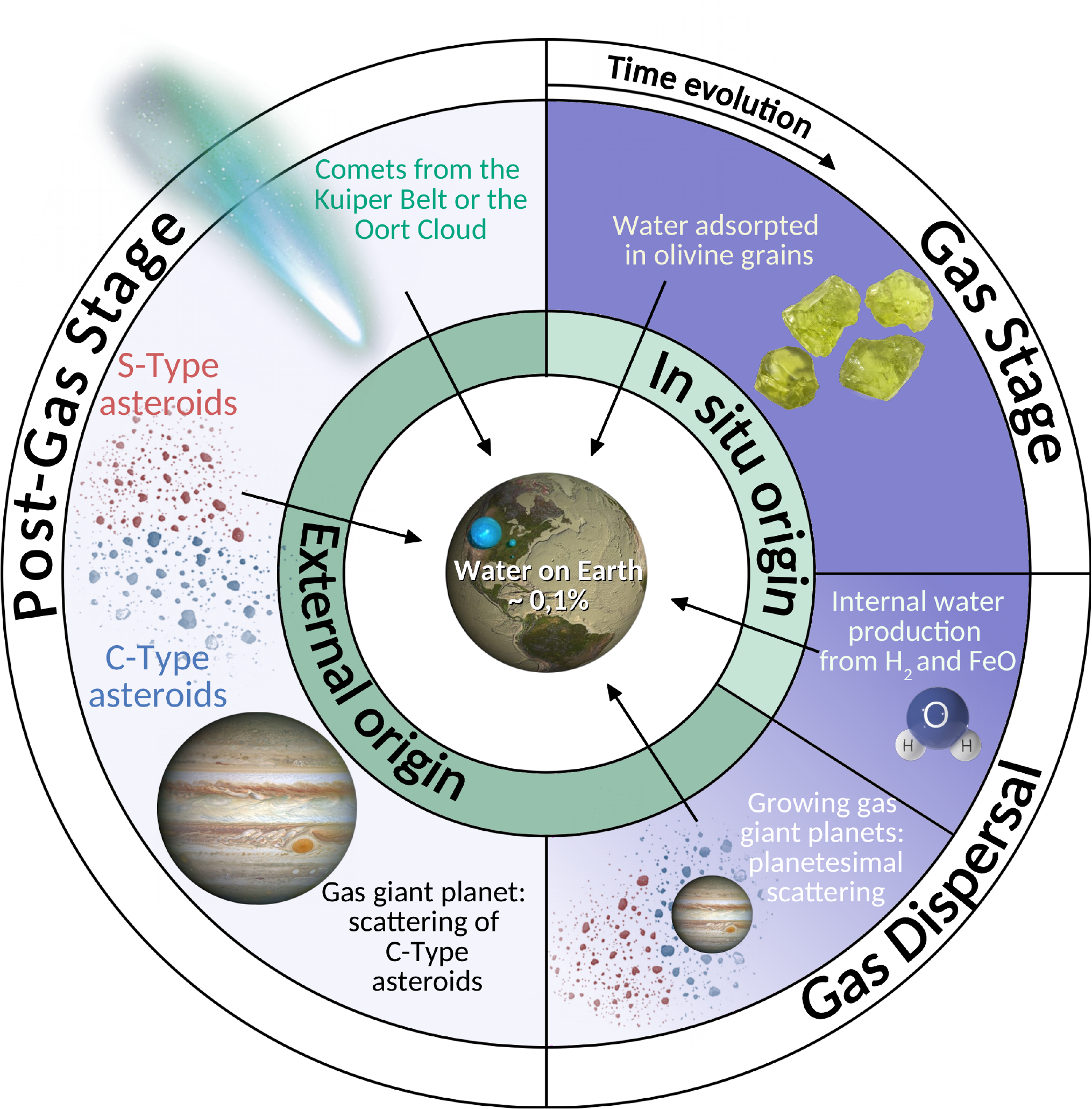}
 \caption{Scheme of the main processes that could contribute to the origin of water on Earth. Central dry Earth illustration by Jack Cook (Woods Hole Oceanographic Institution).}
 \label{fig-water-scheme}
\end{figure}

In the following section, we explain how numerical studies give support to the hypothesis of water being delivered to Earth mainly by C-type asteroids (the parent bodies of carbonaceous chondrites). 
We summarise the main hypothesis of Earth's water origin with a sketch in Fig.~\ref{fig-water-scheme}.

\subsection{Volatile delivery to the terrestrial planets}
\label{Sect:Volatiles2}

Several studies involving numerical simulations show that when the gas disc dissipates, the terrestrial planets form on a timescale ranging from $\sim$30 to 100 Myrs\footnote{This is slightly in tension with Hf-W dating, which shows that the core formation of terrestrial planets occurred at times $\lesssim$30 Myr from the beginning of the Solar System \citep{Kleine02}}, mainly due to giant impacts between embryos and planetesimals distributed along the inner regions of the disc \citep{Wetherill91, Levison2003, Raymond2004, OBrien2006, Kokubo2006, Raymond2006, Morishima2008, Haghighipour2016, Ronco2018}. The gas giants, located farther out and already formed, act as dynamical perturbers that excite the bodies of the main belt, dispersing/scattering water-rich bodies of the outer belt towards the inner regions of the disc where terrestrial planet formation takes place. 
\citet{Morbidelli2000} showed that the fraction of water on Earth could be justified by this mechanism, and other works on the evolution of the Solar System support this idea \citep{OBrien2006, Raymond2006, Raymond2009, OBrien2014, RaymondIzidoro2017}. Additionally, \citet{Izidoro2013} accounts for Earth's water by a combination of embryo/planetesimal accretion and in-situ water production via adsorption of olivine grains. 
We note, however, that the mentioned works are based on the assumption of perfect merging that sets an upper limit for the water content of rocky planets. When a more realistic collision treatment is considered \citep{Chambers2013, Quintana2016}, the terrestrial planets take longer to form ($\sim 100-200$ Myr) and their water content can be reduced at least by a factor of two \citep[][]{Burger2019}.

The classical simulations of terrestrial planet formation and volatile delivery mentioned above, suffered from the inconvenience of not being able to reproduce the low mass of Mars \citep{Wetherill91,Raymond2009}. 
It is worth mentioning that Mars is probably a true embryo from the gas-disc phase, as suggested by radiometric dating with Hf-W isotopes \citep{Dauphas11}. In this regard, the problem of forming a small Mars reduces, simply, to halt its accretion once the disc dissipates. 
The first idea to solve the small-Mars problem is indeed linked to a starving Mars scenario: \citet{Hansen2009} proposed that if the solids in the inner Solar System were concentrated in a ring between 0.7 and 1 AU after the disc's dispersal, Mars could be scattered outwards to a region depleted of material and preserve, in this way, its original mass. 
However, he provided no justification for the existence of that particular ring of solids. Later works showed that such a ring can be an output of planetesimal formation \citep{Drazkowska16}, or the sculpting through migration by gas giants.
This latter option is the one adopted by the \textit{Grand Tack} scenario \citep{Walsh11}. 

The general idea of the Grand Tack is the following. Jupiter, which is already formed, and Saturn, which is still growing, are yet immersed in the protosolar nebula. A dry planetesimal population representing the S-type asteroids remains in the inner regions of the disc, and a water-rich planetesimal population representing the C-type asteroids remains beyond Saturn. Due to its mass, Jupiter is able to open a gap in the gaseous disc and migrate inwards through type II migration. Then, once Saturn approximately reaches its current mass, it migrates inwards faster than Jupiter and gets trapped with it in the 3:2 resonance. While this happens, S-type asteroids are scattered to the outer regions of the disc. If the resonance capture occurs when Jupiter reaches $\sim$ 1.5~au \citep{Walsh11, Brasser2016, Brasser2017}, the inner disc of solids is truncated at $\sim$ 1~au. This scenario reproduces the depleted-disc proposed by \citet{Hansen2009}, necessary for
an already formed embryo scattered from the inner region to the orbit of Mars to remain small.
 \citet{MassetSnellgrove2001} showed that in this configuration of Jupiter and Saturn sharing a common gap, the inward migration can be reversed, making both giants to migrate outwards (see Sect.\ref{Sect:Mig}). While this happens, both planets scatter S-type and C-type asteroids inwards, repopulating the asteroid belt with original S-type asteroids from this region and also with C-type asteroids coming from outer zones. 
The water-rich planetesimals scattered to the terrestrial planet region due to the inwards- then outwards- migration of the giant planets is enough to pollute the Earth. \citet{OBrien2014} analyzed the water delivery to Earth in the Grand Tack context and showed that after the disc dispersal, a fraction of the scattered C-type asteroids could reach the Earth and could explain its current water content. 

The Grand Tack is therefore successful in explaining the volatile content of Earth, the low mass of Mars, and the low-density and composition dichotomy of the current asteroid belt (inner S-type and outer C-type asteroids). It also provides the initial conditions for the latest version of the \textit{Nice Model} \citep{Morbi2007}. The Nice Model \citep{Tsiganis2005,Gomes2005,Morbi2005} proposes that the Solar System underwent a dynamical instability phase after the disc's dispersal, able to explain the current orbits of Jupiter and Saturn, the orbital structure of the Kuiper belt, the orbital distribution of Jupiter's trojan asteroids, and the capture of the irregular satellites of the giant planets \citep[see][and references therein]{Nesvorny2018}. In the original version, the gas giants lie initially on a compact, circular and coplanar configuration, while in posterior versions they start on resonant orbits that emerged during the gas phase \citep{Morbi2007}, as the Grand Tack model provides.
The Grand Tack can be criticised, among other things, for its certainly fine-tuned initial conditions and for neglecting the gas accretion of the gas giants, which can also affect their migration. Some studies have pointed out the low probability of the Grand Tack to occur \citep{Dangelo12, Chametla2020}. For instance, \citet{Chametla2020} find that the capture of Saturn and Jupiter into the 3:2 resonance, and their consequent outwards migration, strongly depends on the initial separation between Jupiter and Saturn. We note, however, that since the Solar System is at the moment unique, low probability scenarios are not a strong argument to discard its plausibility.

Nevertheless, some alternatives to the Grand Tack exists. \citet{RaymondIzidoro2017} propose a different mechanism to pollute the inner regions of the disc with water-rich bodies before terrestrial planets' main growth. The authors show that the growth of the gas giants (with or without migration) naturally excites the eccentricities of water-rich planetesimals of external regions, scattering them in all directions, particularly towards the outer asteroid main belt. The gas, which is still present, damps the eccentricities, allowing them to reach stable orbits in this region. Once the gas dissipates, the planetesimals are no longer damped and some of them reach eccentricities high enough to cross the terrestrial planet formation zone, where they can be accreted by the growing planets. In addition, \citet{RaymondIzidoro2017b} show that even if the asteroid belt was empty from the beginning, as some planetesimal formation models proposed \citep[e.g.][]{Drazkowska16}, this model can still reproduce the current mass and dichotomy of the asteroid belt. This is a natural byproduct of, on one hand, the S- and C-type planetesimal implantation in the asteroid belt region due to the growing gas giants, and on the other, the planetesimal scattering from the terrestrial planet formation zone. It is important to remark that, within this new scenario, the formation of a small Mars can also be achieved.

\subsection{Why is the Earth so dry?: the role of Jupiter}
\label{Sect:Jupiter}

Despite that it is widely accepted that embryos and planetesimals remained in the disc after its dispersal, the dominant size of the planetary building blocks during the gas phase has been recently re-addressed, as we explained in Sect.\ref{Sect:Sizes}. Certainly, pebble-based formation models have gained popularity, mainly due to its ability to explain a fast planet formation \citep{JohansenLambrechts2017}, which could account for some of the ring-structures observed with ALMA in young discs \citep{Lodato2019, Ndugu2019}.
However, if pebbles are the main responsible of building embryos and planets within the disc, this poses a very different picture for the origin of volatiles compared to the classical scenario described above. A major problem that arises, as we mentioned in Sect.\ref{Sect:Volatiles}, is how to leave the inner regions of a planetary system dry, since icy pebbles drift and very likely reach the inner regions as the discs cools, polluting it with volatiles \citep{Bitsch19,Ida2019}. Avoiding such a problem in the Solar System is probably related to the early growth of Jupiter \citep{Morby15}.

Meteoritic record gives support to this idea. Recent re-analysis of meteoritic data shows that carbonaceous and non-carbonaceous chondrites (hereafter CC and NC, respectively) have very distinct Tungsten isotopic anomalies within iron meteorites \citep{Kruijer17}. 
These differences imply a different accretion time for the two types of reservoirs. 
In particular, it implies that after time $\sim$1 Myr from the formation of CAIs  (Calcium-Aluminium-rich Inclusions), the two groups were spatially separated. 
From the dating it is also inferred that CC finished their accretion at times $\sim$3-4 Myr after CAI \citep{Kruijer17}. 
Consequently, \citet{Kruijer17} conclude that the two reservoirs were separated from each other between times $\sim$1 and $\sim$3-4 Myr after CAI, and reconnected afterwards. 
The authors claim that the best explanation for this is the formation of Jupiter: the core of the giant planet formed early, and the growing planet acted as a barrier that prevented material from mixing during $\sim$2-3 Myr. 
 More specifically, Jupiter acquired its pebble isolation mass (see Sect. \ref{Sect:Sizes}) at time $\sim$1 Myr after CAI, preventing icy pebbles from beyond Jupiter's orbit to reach the region within it after this time. 
Then, in the next $\sim$2 Myr, NC and CC accreted separated from each other at both sides of the planet, and finally, when the planet reached approximately 50 \ME, it was massive enough to scatter these solids, reconnecting the reservoirs and probably scattering water-rich planetesimals to the forming rocky planets \citep{RaymondIzidoro2017}.

\citet{A18} built upon the \citet{Kruijer17} scenario and showed that for it to occur, Jupiter had to grow by an initial phase of pebble accretion, followed by planetesimal accretion. 
The idea is the following. After Jupiter reaches the pebble isolation mass (which provokes the separation of the NC and CC reservoirs), pebble accretion stops by definition (see Sect.\ref{Sect:Iso}). 
During the growth of a gas giant, if solid accretion is halted when the protoplanet is close or beyond critical mass\footnote{A typical value of the pebble isolation mass at $a$= 5 au is $\sim$20 \ME \citep{Lambrechts14}, which is slightly larger than typical values of the critical core mass \citep{Ikoma00}.}, then runaway gas accretion proceeds extremely fast \citep{Ikoma00}. The reason for this is that the loss of heating from solid accretion decreases drastically the envelope's pressure support, making the envelope more prone to contract due to self-gravity. As a consequence, once the pebble isolation mass is reached, and without any additional source of heating, the formation of Jupiter after reaching pebble isolation mass would last only $\sim0.1$ Myr  \citep{Ikoma00,A18}. 
This is in clear contradiction with the timescale inferred by \citet{Kruijer17} from the cosmochemical data. \citet{A18} found that in order to fulfill the meteoritic constraints, an extra source of solid accretion is necessary to make Jupiter grow from pebble isolation mass until $\sim$50 \ME \, in a time lapse of $\sim$2-3 Myr. Since this extra source of solids cannot be pebbles, it must necessarily be planetesimals. Furthermore, \citet{Venturini20} showed that under this hybrid pebble-planetesimal accretion scenario, the metallicity of Jupiter \citep{Wahl17} is naturally explained.

The scenario proposed by \citet{A18} might also shed light into the problem of the dominant size of solids in planet formation models. Perhaps it is not pebbles \textit{versus} planetesimals, but a dominance of pebble accretion at the earliest stages to make the core grow fast, followed by a later accretion of planetesimals to delay gas accretion for a few million years. Hence, this hybrid model might also help to understand the abundance of intermediate-mass exoplanets.

\section{Exoplanets and volatile content}
\label{Exoplanets}

From exoplanets we have much less data than for Solar System objects.
For approximately $\sim$4000 exoplanets the radius is known, from which only $\sim$100 have determined masses below 25 \ME \,\citep{Lozovsky18}. Only for few exoplanets we have some hint of the atmospheric composition from transmission spectroscopy \citep{Kreidberg2017}.
The \textit{Kepler} mission revealed that one of the most common type of exoplanets are those with a radius between that of Earth and Neptune \citep{Batalha13}. These planets do not exist in our Solar System, and could in principle be larger versions of our rocky Earth (\textit{super-Earths}) or smaller versions of the ice-giants (\textit{mini-Neptunes}). 
A second important surprise revealed by Kepler was the finding of a bi-modal distribution in planet sizes
for exoplanets with orbital period of less than 100 days, with a peak at 1.3 and 2.4 Earth radius \citep{Fulton17}. While the gap between the two peaks (at about 1.8 Earth radius) could be partially filled by unseen close stellar companions \citep{Teske2018}, this effect cannot account for the persistent bi-modality.

Theoretical models that include photoevaporation due to the central star can match this bi-modal behaviour \citep{Lopez12, Owen13, Owen17, JInMord18}, which made the dearth of the radii distribution to be known as the \textit{photoevaporation valley}. Note, however, that an alternative explanation for the bi-modality in planetary radii exists, known as \textit{core-powered mass-loss} \citep{Ginzburg18, Gupta19}. In this scenario, the heat remaining from formation strips off tenuous atmospheres once the disc dissipates. In addition, the \textit{Parker wind} mechanism could also account for the mass-loss of planetary envelopes at the time of disc dissipation \citep{Owen2016}. This mechanism requires the radius of the planet to be approximately the Bondi radius when the disc disappears. However, \citet{Bodenheimer2018} show, in recent planet formation simulations of close-in super Earths that account for the thermal structure of a silicon-rich envelope, that the planet radius at such time should be only $\sim$10\% of the Bondi radius.

Why do photoevaporation and core-powered mass-loss produce two peaks in the radii distribution? In the case of photoevaporation, if a planet made by a rocky core and surrounded by a very thin H-He atmosphere (less than 1\% of the planet's mass) is exposed to X-ray and EUV irradiation, the hydrogen and helium acquire a thermal speed larger than the escape speed of the planet. This triggers hydrodynamical escape, and after some giga-years of irradiation exposure, the planet is left solely as a naked rocky core \citep{Lopez12, Johnstone2015}. Pure rocky planets cannot have radius larger than 1.6 \citep{Rogers15}. Thus, in the view of photoevaporation models, the first peak of the radii distribution corresponds to naked rocky cores, while the planets from the second peak should posses some gaseous envelope.
Indeed, the work of \citet{Owen17} shows that the timescale to lose the envelope by photoevaporation is the longest if the envelope represents $\sim$1-10\% of the planet's mass. Hence, planets with such atmospheres are more stable against photoevaporation and would constitute the second peak of the distribution found by \citet{Fulton17}.  
Similarly, in the case of core-powered mass-loss, the heat of the core strips atmospheres if they are less massive than $M_{\rm atm}/M_{\rm core} \sim 5\% $ \citep{Ginzburg18}. 
Atmospheres with masses above this threshold will compactify, increasing their binding energy and making mass-loss more difficult, while atmospheres with initial mass below the threshold will expand, and once the process is triggered, the smaller is the binding energy, and the easier it is for the remaining atmosphere to be removed.
 
An important aspect of both scenarios is that the position of the valley is extremely sensitive to the composition of the core. The models only match the observations if the cores are assumed to be rocky \citep[with a maximum ice mass fraction of $\sim$ 20\% in the case of core-powered mass-loss,][]{Gupta19}.
This has lead to the conclusion that the bulk of these exoplanets are dry, and therefore, formed inside the water ice-line. 
It is worth mentioning that these models assume always an envelope made purely by hydrogen and helium. 
A work by \citet{Kurosaki14} shows that planets with pure water envelopes with radii less than $\sim$2-3 Earth radius and larger than that of a pure rocky composition, would survive photoevaporation. Hence, small mass planets (M$\lesssim$10 \ME) with rocky cores and water envelopes could contribute to populate the non-empty valley and the second peak of the distribution. In the same line of thought, based on mass-radius relations, \citet{Zeng19} argue that planets composed of 50\% rock and 50\% water by mass could populate the second peak of the distribution.

If most short period exoplanets are indeed depleted of water as photoevaporation and core-powered mass-loss models suggest, this poses a problem for formation models, 
which tend to show that planets with non-negligible H-He atmospheres contain large portions of water \citep{Alibert13, Venturini17}. More precisely, the problem is two-fold: first, to have enough rocky material at short orbital distances to form planets larger than Earth able to bind some significant H-He envelope; second, to avoid that such planets enter in the runaway gas phase and become gas giants.

In the case of pebble accretion the problem is that, without the presence of a giant planet in an outer orbit, icy pebbles can reach the inner parts of the disc, polluting formed planets with water, as mentioned in Sect.\ref{Sect:Volatiles2}. Even if the iceline does not reach the regions of orbital period shorter than 100 days \citep{Bitsch19}, planets in the mass range of $\sim$ 5-15 Earth masses \citep[the ones that would populate the second peak of the Kepler radii distribution,][]{Zeng19} tend to migrate efficiently. Thus, they could form beyond the iceline, accrete considerable amounts of ice, and then migrate and park at orbits within a 100-day period \citep{Izidoro19,Bitsch19}. 
Still, rocky super-Earths could be produced by silicate pebble accretion inside the iceline if the pebble flux exceeds certain threshold \citep{Lambrechts19}.
Alternatively, N-body simulations have shown that super-Earths can form as a result of collisions of rocky embryos \citep{Ogihara18a,Raymond2018}, and that runaway of gas could be halted due to the limited gas supply from a viscous disc \citep{Ogihara18b}.
Even if these two last mechanisms could produce rocky super-Earths efficiently, it is not totally clear how icy migrating planets can be prevented from finishing at the same location, which would smear out the valley of the distribution \citep{VanEylen18}. \citet{Ogihara18a} propose that disc winds flatten the gas disc's profile, abating type I migration and hence the presence of icy planets in the inner disc. This interesting scenario deserves further development, in particular because the predominance of disc winds is needed to halt type-I migration \citep{Ogihara18a}, but it is not enough to prevent runaway of gas onto the formed super-Earths \citep{Ogihara18b, Ogihara2020}.

In the case of pure planetesimal accretion, studies that attempted to form super-Earths inside the ice-line show that this is difficult \citep{IkomaHori12, BodLiss14}.
The main reason for this is that there is typically not enough solid material inside the ice-line to form planets able to bind a substantial H-He envelope, and also, gas accretion in hot regions of the disc is less effective. 
Recent population synthesis with planetesimal accretion show that dry super-Earths can form inside the water ice-line, but water-rich migrating planets also arrive at short orbital distances at the time of disc dispersal \citep[Fig.8 of][]{Mordasini2018}. Indeed, there is a trade-off between the disc mass and the water content of short-period exoplanets: more massive discs are able to form larger dry super-Earths, but, at the same time, form larger icy objects which migrate to the inner regions more efficiently\footnote{Type-I migration is faster for more massive planets, see Sect.\ref{Sect:Mig}.} \citep{Mordasini2018}. 

Regarding the problem of avoiding the runaway of gas onto the super-Earths, in principle envelope enrichment (Sect.\ref{Sect:CAM}) complicates the matter even more. Especially when the size of the accreting solids is small (i.e, pebbles and small planetesimals, which would not create strong compositional gradients), pollution by the sublimated incoming solids into the H-He atmospheres of growing planets must certainly happen \citep{Brouwers2018}. This enhances gas accretion, making the runaway of gas more likely to happen during the disc's lifetime \citep{Venturini16}. A mechanism proposed to halt gas accretion is the \textit{recycling} of atmospheric gas into the disc \citep{Ormel15,LambrechtsLega17,Cimerman2017}. However, this phenomenon has been recently challenged by  hydrodynamical simulations that account for radiative transfer via a beta-cooling approximation \citep{Kurokawa18}.
Further theoretical work that includes different physical aspects into a commom framework; together with a broader determination of planetary masses, will help to elucidate the composition and formation paths of exoplanets.

\section{Summary and conclusions}
\label{Sect:Summary}

This review aimed at summarising, in a nutshell, key concepts of planet formation such as critical core mass, pebble isolation mass, migration types, icelines and volatile delivery; 
as well as providing an overview of the main processes and open questions in the field.

Regarding planet formation theory, it is becoming widely accepted the major role of streaming instability on forming the planetesimals that serve as seeds for core accretion \citep{Youdin05,Johansen07,Drazkowska16}.
Also increasingly accepted is the fact that particle traps like pressure maxima, and migration convergent zones, are preferential locations for planet formation and survival \citep{Dittkrist2014, Baillie2016, CL2016, Guilera2017, Cridland2017, Ndugu2018, Pudritz2018}. Although still numerically challenging, solid and gas accretion onto protoplanets has to be computed together with planet migration, because the timescales of these processes are comparable.

Numerous important processes in planet formation have emerged in the last years, like pebble accretion \citep{OrmelK10, Lambrechts12}, gas recycling \citep{Ormel15}, envelope enrichment \citep{HI11, Venturini15, Venturini16, Brouwers2018}, corotation and thermal torques \citep{Paardekooper2011, Lega2014, Benitez-llambay2015, jm2017, masset2017}. 
Each of them has to be included in global simulations to asses the effect on the final output of formation. 
In parallel, new constraints for planet formation are rising from ALMA observations of disc structure and composition \citep{DSHARP1}.
 In particular, the presence of rings in discs observed by ALMA is highlighting the importance that these features might have on the outcome of planet formation \citep{Morby2020}.
 
Regarding the volatile content of the Solar System (and Earth in particular), we can summarise the state-of-the-art knowledge as follows:

\begin{itemize}
\item Earth and Venus were very likely able to bind some primordial H$_2$ atmosphere by the time of disc dispersal. Geochemical evidence for this stands from the isotopic records of noble gases \citep{Marty2012} and from the low D/H values found in terrestrial lavas \citep{Hallis2015}. For that primordial H$_2$ atmosphere to exist and later dissipate, formation and atmospheric escape models constrain the mass of Venus and Earth at the time of disc dispersal to be in the range of $\sim$0.5-0.75 \ME \, \citep{Lammer2018}.
\item Mars was probably fully formed by the time of disc dissipation \citep{Dauphas11}, which might have happened at time $\sim$4 Myr after CAIs formation \citep{Wang2017}.
\item D/H and ${}^{15}$N/${}^{14}$N ratios on Earth's crust match those of carbonaceous chondrites \citep{Marty2012}. 
\item Water delivery to Earth by comets can account for very little ($\lesssim$1\%) of its water, but comets might have contributed greatly to the budget of Earth's noble gases \citep{Marty2016}.
\item Overall, Earth formation took $\sim$30-100 Myr \citep[e.g,][]{Izidoro18}, and water delivery during the disc phase would have produced a too high oxidation state \citep{Rubie2011}.
\end{itemize}
The last three points strengthen the view that volatiles (and particularly water) on Earth were delivered mainly from chondritic material (asteroids and not comets) during and/or after the disc dispersal \citep{Marty2012, OBrien18}. 
Still, the exact contribution of water from nebular origin is hard to determine, given the poor knowledge on the exact amount and isotopic composition of water in the terrestrial mantle and core. 

Also linked to the volatile content of terrestrial planets, it is widely accepted that Jupiter acted as a barrier that prevented carbonaceous chondritic and water-rich material from reaching $a \sim 1$ au during the gas disc phase \citep{Morby15, Kruijer17}.
The timing of this event gives support to Jupiter being formed in a hybrid pebble-planetesimal fashion \citep{A18, Venturini20}.
Both cosmochemical data and formation theory point towards the major role of Jupiter in keeping the inner Solar System dry.

Regarding exoplanets, despite of possessing much less information than with Solar System objects, a combination of radii determination and thermal evolution models suggests that exoplanets with periods less than $\sim$100 days and radius smaller than that of Neptune are water poor. 
This poses a problem for planet formation models, which tend to predict too much water within super-Earths/mini-Neptunes.
Possible answers to this problem might come from missing physical processes in the models. For instance, \citet{Lichtenberg19} recently showed that when the heating from radioactive decay is included in planet formation with planetesimal accretion, the final outcome is much drier planets.
Water might also be lost during planet evolution, for example due to the remanent heat from the core \citep{Vazan18}. 

Better radii determination will come with the ongoing missions of TESS and  CHEOPS and with near-future missions like PLATO. Furthermore, a wealth of exoplanet atmospheric spectra will be acquired by JWST and ARIEL. 
More data coupled with more physically motivated models will be crucial to unveil the composition of exoplanets.

\section*{Acknowledgements} We thank the International Space Science Institute for organising the workshop "Understanding the Diversity of Planetary Atmospheres" and the workshop participants for the encouragement to write this review/chapter. We also thank the Editor, Dr. Helmut Lammer, and the anonymous referees for their valuable comments and suggestions which helped us to significantly improve this work. MPR acknowledges financial support provided by FONDECYT grant 3190336 and from CONICYT project Basal AFB-170002. MPR and OMG acknowledge financial support from the Iniciativa Cient\'{\i}fica Milenio (ICM) via the N\'ucleo Milenio de Formaci\'on Planetaria Grant. OMG is partially supported by the PICT 2016-0053 from ANPCyT, Argentina. OMG acknowledges the hosting by IA-PUC as an invited researcher.

\bibliographystyle{spbasic}

\bibliography{lit_2019}

\end{document}